\documentclass[11pt,twocolumn,prb,superscriptaddress,reprint]{revtex4-1}

\usepackage{graphicx,amsmath,amssymb,color}
\usepackage{tabularx, ctable}

\usepackage{ulem}
\usepackage{bm}
\usepackage{siunitx}
\usepackage{gensymb}
\usepackage{xr}
\usepackage[breaklinks=true,colorlinks,allcolors=blue]{hyperref}

\newcommand{\be}{\begin{eqnarray}}
\newcommand{\ee}{\end{eqnarray}}

\begin{document}

\title{Enhanced Terahertz Thermoelectricity via Engineered van Hove Singularities and Nernst Effect in Moiré Superlattices}

\author{L. Elesin$^{+}$}

\affiliation{Department of Materials Science and Engineering, National University of Singapore, 117575, Singapore}

\author{A. L. Shilov$^{+}$}
\affiliation{Department of Materials Science and Engineering, National University of Singapore, 117575, Singapore}

\author{S. Jana$^{+}$}
\affiliation{Department of Materials Science and Engineering, National University of Singapore, 117575, Singapore}

\author{I. Mazurenko}
\affiliation{Programmable Functional Materials Lab, Center for Neurophysics and Neuromorphic Technologies, Moscow, 127495}

\author{P. A. Pantaleon}
\affiliation{Imdea Nanoscience, Faraday 9, 28015 Madrid, Spain}

\author{M.~Kashchenko}
\affiliation{Programmable Functional Materials Lab, Center for Neurophysics and Neuromorphic Technologies, Moscow, 127495}
\affiliation{Center for Advanced Studies, Kulakova str, Moscow}

\author{N. Krivovichev}
\affiliation{Programmable Functional Materials Lab, Center for Neurophysics and Neuromorphic Technologies, Moscow, 127495}

\author{V.~Dremov}
\affiliation{Programmable Functional Materials Lab, Center for Neurophysics and Neuromorphic Technologies, Moscow, 127495}

\author{I. Gayduchenko}
\affiliation{Moscow Pedagogical State University, Moscow 119991}

\author{G. Goltsman}
\affiliation{Moscow Pedagogical State University, Moscow 119991}

\author{T. Taniguchi}
\affiliation{International Center for Materials Nanoarchitectonics, National Institute of Material Science, Tsukuba 305-0044, Japan}

\author{K. Watanabe}
\affiliation{Research Center for Functional Materials, National Institute of Material}

\author{Y. Wang}
\affiliation{Institute for Functional Intelligent
Materials, National University of Singapore, Singapore, 117575, Singapore}

\author{E. I. Titova}
\affiliation{Programmable Functional Materials Lab, Center for Neurophysics and Neuromorphic Technologies, Moscow, 127495}
\affiliation{Center for Advanced Studies, Kulakova str, Moscow}

\author{D. A. Svintsov}
\affiliation{Center for Advanced Studies, Kulakova str, Moscow}

\author{K. S. Novoselov}
\affiliation{Institute for Functional Intelligent
Materials, National University of Singapore, Singapore, 117575, Singapore}

\author{D. A. Bandurin}
\affiliation{Department of Materials Science and Engineering, National University of Singapore, 117575, Singapore}
\affiliation{Institute for Functional Intelligent
Materials, National University of Singapore, Singapore, 117575, Singapore}

\begin{abstract}

\textbf{Thermoelectric materials, long explored for energy harvesting and thermal sensing, convert heat directly into electrical signals. Extending their application to the terahertz (THz) frequency range opens opportunities for low-noise, bias-free THz detection, yet conventional thermoelectrics lack the sensitivity required for practical devices. Thermoelectric coefficients can be strongly enhanced near van Hove singularities (VHS), though these are usually difficult to access in conventional materials. 
Here we show that moiré band engineering unlocks these singularities for THz optoelectronics.
Using 2D moiré structures as a model system, we observe strong enhancement of the THz photothermoelectric response in monolayer and bilayer graphene superlattices when the Fermi level is tuned to band singularities. Applying a relatively small magnetic field further boosts the response through the THz-driven Nernst effect, a transverse thermoelectric current driven by the THz-induced temperature gradient. Our results establish moiré superlattices as a versatile platform for THz thermoelectricity and highlight engineered band structures as a route to high-performance THz optoelectronic devices.}




\end{abstract}

\maketitle

\begin{figure*}[ht!]
  \centering\includegraphics[width=\linewidth]{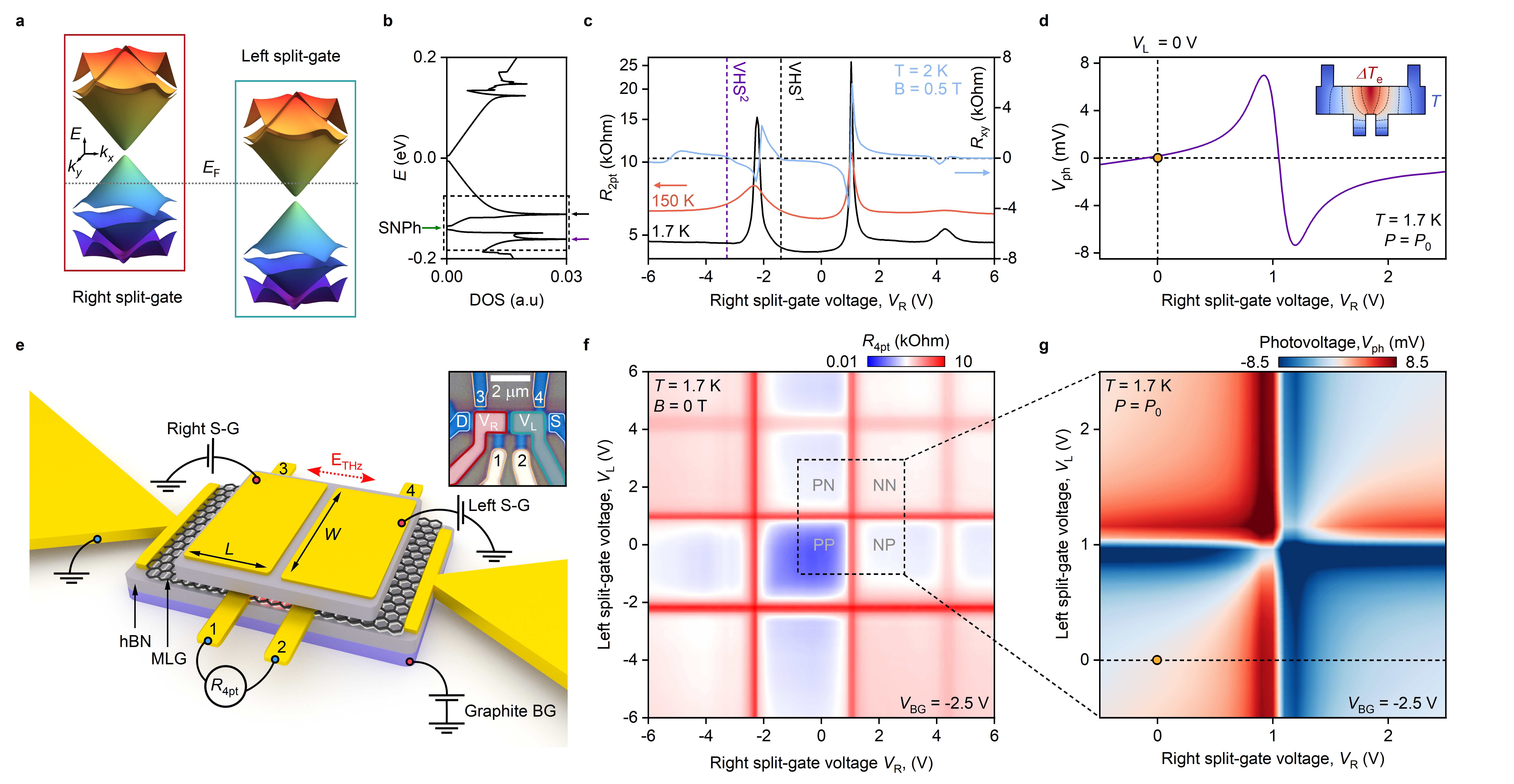}
    \caption{\textbf{Transport characteristics and PTE effect in the split-gate MLG/hBN MS.}
    \textbf{a,} Schematic of the band structure in two regions of the split-gate junction. The left and right split-gates allow independent control of the carrier type and Fermi level in the junction segments. \textbf{b,} DOS of the perfectly-aligned MLG/hBN MS. \textbf{c,} Magnetotransport measurements of the two-probe resistance as a function of $V_\mathrm{R}$ at 1.7 K (black) and 150 K (red). The blue line shows Hall resistivity $R_{xy}$ of the right split-gate segment, obtained by antisymmetrization at $B = \pm 0.5$~T. Vertical dashed lines mark the position of VHSs. \textbf{d,} Individual $V_{\mathrm{ph}}$ trace as a function of the right split-gate voltage $V_{\mathrm{R}}$ at a fixed left split-gate voltage $V_{\mathrm{L}} = 0$. The vertical dashed line corresponds to the condition $V_{\mathrm{L}} =V_{\mathrm{R}}$ where zero net photovoltage is expected.  \textbf{e,} Split-gate device schematic showing antennas connected to the source and drain terminals and aligned with THz polarization. Inset: optical photograph of the MLG/hBN MS sample. \textbf{f,} Dual-gate map of the four-probe resistance $R_{\mathrm{4pt}}$ of the junction at a fixed back-gate voltage $V_{\mathrm{BG}}$ = -2.5 V. \textbf{g,} Dual-gate map of photovoltage $V_{\mathrm{ph}}$ near the charge neutrality point. $V_\mathrm{ph}$ profile across the dashed line is shown in (d).
}
	\label{Fig1}
\end{figure*}

Millimeter-wave and terahertz (THz) radiation are increasingly important for applications ranging from security screening and ultrafast wireless communications to medical imaging and radio astronomy\cite{saeedkia2013handbook, leitenstorfer20232023}. Efficient detection, however, remains technologically challenging and is ultimately constrained by the intrinsic electronic properties of the materials employed. Conventional semiconductor devices, such as GaAs-based Schottky diodes~\cite{sizov2010thz, bulcha2016design, mehdi2017thz}, offer fast, room-temperature detection, but their responsivity degrades rapidly beyond $1$~THz and they operate only as incoherent receivers. Superconducting hot-electron bolometers~\cite{klapwijk2017engineering, shurakov2015superconducting, shein2024fundamental} (SHEBs), on the other hand, exhibit outstanding responsivity and minimal noise equivalent power (NEP) but require liquid helium temperatures. Field-effect transistor (FET)-based THz sensors have emerged as a promising alternative\cite{dyakonov2002detection, shur2005terahertz, tauk2006plasma, knap2009field}, combining broadband response, plasmonic enhancement~\cite{dyakonov2002detection} and the possibility of phase sensitive detection\cite{glaab2010terahertz, rumyantsev2017homodyne} for noise-immune communications yet their performance is fundamentally limited by carrier mobilities and plasmon lifetimes, both sensitive to material quality \cite{muravev2016response}. The path forward thus lies in emergent material platforms with complex band structures and tunable properties that can overcome the intrinsic trade-offs of traditional materials~\cite{SnTe}.

Graphene has emerged as a promising alternative for this inquiry, owing to its exceptional electronic properties which has given rise to a plethora of operative THz detection mechanisms, including resistive self-mixing \cite{vicarelli2012graphene, spirito2014high}, plasmon-assisted detection\cite{bandurin2018resonant, caridad2024room}, tunnel\cite{gayduchenko2021tunnel} and Schottky-type~\cite{schlecht2019efficient} nonlinearities, hot electron \cite{wei2008ultrasensitive, lara2019towards} and viscous bolometry\cite{kravtsov2025viscous}, ballistic rectification\cite{auton2017terahertz, do2025nonlinear}, and photothermoelectricity \cite{xu2010photo, cai2014sensitive, bandurin2018dual, castilla2019fast, titova2023ultralow}, to name a few.
Among them, the latter stands out for its zero-bias broadband operation, strong responsivity, and low NEP and ultrafast response\cite{castilla2019fast,titova2023ultralow, soundarapandian2024high}. This is because high tunability of graphene's Fermi level via electrostatic gating enables engineering of p-n junctions in the split-gate geometries\cite{castilla2019fast, castilla2020plasmonic}, where absorbed THz power, that causes an increase in electron temperature in the middle of the junction, can be readily converted into a strong DC thermoelectric photovoltage. The latter is governed by the thermoelectric tensor that, within the semiclassical formalism\cite{cutler1969observation}, scales with the derivatives of the conductivity tensor $\sigma$ components over carrier density $n$ and the density of states (DOS) function $D(E)$ at the Fermi level $E_\mathrm{F}$:

\begin{equation}
S_{ij} = -\frac{\pi^2 k_B^2 T}{3 |e|} D(E_\mathrm{F})\sum\limits_k
(\sigma^{-1})_{ik} \frac{d \sigma_{kj}}{d n},\label{eq1}
\end{equation}
where $k_\mathrm{B}$ is the Boltzmann constant, $e$ is the elementary charge, and $T$ is the temperature. 
All graphene detectors studied so far have operated in the Seebeck regime at zero magnetic field, where $d\sigma_\mathrm{xx}/dn$ peaking near the charge neutrality point (CNP) is counteracted by the vanishing DOS, ultimately limiting the efficiency of photothermoelectric (PTE) rectification.



In this work, we circumvent this problem by enhancing the PTE response through band structure engineering in monolayer and bilayer graphene-based moir\'e superlattices (MS). When the graphene lattice is aligned with the underlying hexagonal boron nitride (hBN) substrate, the  moir\'e pattern forming on the interface substantially alters the electronic band structure, giving rise to distinct features in $D(E)$ (Fig.~\ref{Fig1}a), including secondary neutrality points (SNP) and VHS~\cite{yankowitz2012emergence, hunt2013massive, ponomarenko2013cloning}. We demonstrate that when the Fermi level in the graphene/hBN MS is tuned close to the hole-side SNP (SNPh), the PTE response under THz illumination is substantially enhanced due to the steeper energy dependence of the $D(E)$ compared to pristine graphene, associated with a reduced Fermi velocity $v_\mathrm{F}$ in moir\'e bands. When the doping is further adjusted to reach a VHS near the SNPh, the PTE rectification is dramatically enhanced in the presence of a magnetic field $B$ due to the strong amplification of the off-diagonal components of the Seebeck tensor and the resulting Nernst photovoltage. These findings establish moir\'e engineering as a viable pathway for advancing the performance of graphene-based THz optoelectronic devices.





\begin{figure*}[ht!]
  \centering\includegraphics[width=1\linewidth]{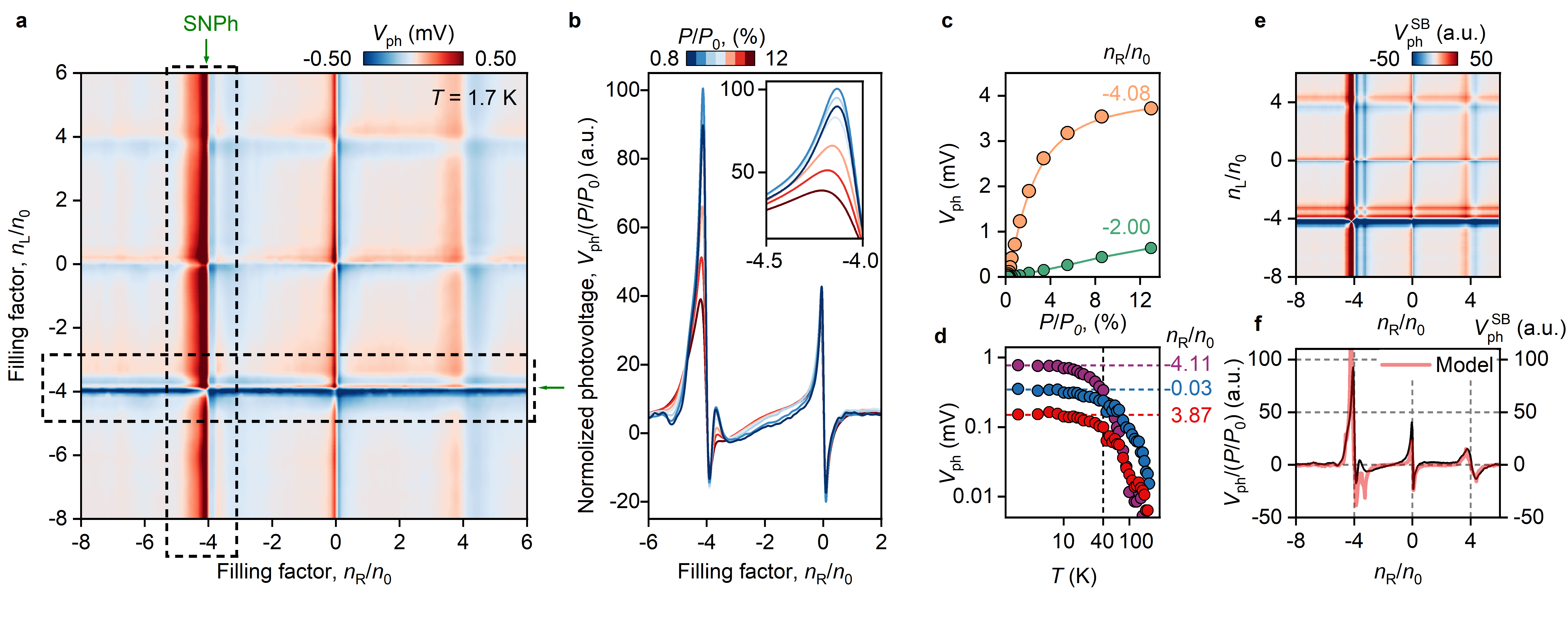}
    \caption{\textbf{Enhancement of the THz-driven Seebeck effect near the SNPh in the MLG/hBN MS.}
    \textbf{a,} Dual-gate photovoltage map at  $P/P_{\mathrm{0}} = 0.08\%$. Dashed rectangles highlight the region near SNPh where the photovoltage is enhanced. \textbf{b,} Normalized photovoltage as a function of the right-side filling factor $n_{\mathrm{R}}/n_0$ under varying attenuated power. Inset: enlarged region near filling factor $n_{\mathrm{R}}/n_0 = -4$. \textbf{c,}  Photovoltage as a function of attenuated power at  selected right-side doping $n_R/n_0$ and fixed $n_L/n_0 = 3$. \textbf{d,} Temperature dependence of photovoltage near NPs at fixed attenuated power $P/P_{\mathrm{0}} = 0.08\%$. \textbf{e,} Theoretical prediction of the dual-gate map based on the Mott formula. \textbf{f,} Comparison between experimental measurements and theoretical modeling of Seebeck photoresponse.
    }
	\label{Fig2}
\end{figure*}

\textbf{Device structure and transport characterization.}
To explore PTE effects in MS, we fabricated devices based on monolayer (MLG) and bilayer graphene (BLG), each encapsulated in two hBN slabs\cite{castellanos2014deterministic}, with one of the slabs crystallographically aligned with the graphene layer. The MLG devices were patterned in the form of multiterminal split-gate FETs, enabling independent control of the carrier density in each channel segment. In addition, one of the MLG samples was equipped with a graphite back gate (see Fig.~\ref{Fig1}e), enabling doping of the whole device (only the data for this MLG device is discussed in the main text; the data for the second MLG device is shown in Supplementary information). Source and drain electrodes in the MLG devices were connected to a broadband antenna that funnels incident THz radiation into the device. In contrast, the BLG sample was patterned into a conventional Hall-bar geometry, with a single top gate and one of the contacts connected to the antenna (see Fig. \ref{Fig4}a). In total, three devices were studied, all exhibiting similar behavior. The samples were mounted in the chamber of the magneto-optical variable-temperature cryostat and illuminated with sub-terahertz radiation ($f_\mathrm{THz}=0.14$ THz) via an external optical setup, comprising lenses and mirrors\cite{shilov2024high}. 


Before turning to photoresponse measurements, we first characterize the transport properties of our MSs. Transport data measured in MLG/hBN MS reveal signatures characteristic of a moiré system, with three conventional peaks in electrostatic doping dependence of the longitudinal resistance (Fig.\ref{Fig1}c). While the middle peak is the remnant of the MLG's original neutrality point, the two side peaks appear at the SNPs arising from  moiré-induced band reconstruction (Fig.\ref{Fig1}b and Supplementary Information). Figure \ref{Fig1}f shows the four-terminal resistance across the junction $R_\mathrm{4pt}$, mapped as a function of the split-gate voltage configurations ($V_\mathrm{L},V_\mathrm{R})$ at a fixed back gate voltage $V_{\mathrm{BG}} = -2.5$V. The resulting chessboard pattern in this map reflects the formation of p-n junctions between regions of opposite carrier types. 


\textbf{THz-driven Seebeck effect in MLG/hBN superlattices}.
To investigate the response under continuous 0.14 THz irradiation, we employed a standard lock-in technique, measuring the photovoltage at the modulation frequency of a THz source\cite{shilov2024high}. Figure \ref{Fig1}d shows the photovoltage $V_\mathrm{ph}$ built up across the device as a function of the right split-gate voltage $V_\mathrm{R}$ in the vicinity of the main neutrality point (NP), with the left split-gate voltage fixed at $V_{\mathrm{L}} = 0$ V. The observed dependence exhibits the characteristic behavior of graphene-based THz detectors \cite{titova2023ultralow, gabor2011hot}, including the conventional sign reversal upon crossing the charge neutrality point at $V_\mathrm{R} = 1$ V, corresponding to a change in carrier type~\cite{PhilipThermo}. However, instead of displaying the expected monotonic decrease of response amplitude far from the main neutrality point, the photovoltage exhibits a second sign reversal at  $V_{\mathrm{R}} = 0$ V. This sign change arises from the split-gate geometry and indicates the point at which the Seebeck coefficients on the left and right sides of the junction become equal. Indeed, the thermoelectric photovoltage $V^{SB}_\mathrm{ph}$ generated across the channel with the split-gate is given by: 
\begin{equation}
\label{eq2}
\begin{split}
V_{\mathrm{ph}}^{SB} 
&= \int_{-L}^{L}  S_{xx}(x) \, \frac{\partial T_{\mathrm{e}}}{\partial x}\, dx \\
&\approx \left[ S^\mathrm{(L)}_{xx} - S^\mathrm{(R)}_{xx} \right] \, \Delta T_\mathrm{e} ,
\end{split}
\end{equation}
where $L$ is the length of each split-gate segment, $S^\mathrm{(L)}_\mathrm{xx}$ and $S^\mathrm{(R)}_{xx}$ are the local Seebeck coefficients in the regions beneath the left and the right split-gates, respectively, and $T_\mathrm{e}$ is the variation of electron temperature across the sample. Due to the  symmetric design of the device, $T_\mathrm{e}$ is maximized in the middle of the split junction and is $\Delta T_\mathrm{e}$ larger than that at the source and drain terminals, which are thermalized with the bath at $T$. 

As a result, $V_{\mathrm{ph}}$ cancels out when $S^\mathrm{(L)}_{xx}=S^\mathrm{(R)}_{xx}$.
Figure \ref{Fig1}g shows $V_{\mathrm{ph}}$ plotted as a function of $V_\mathrm{L}$ and $V_\mathrm{R}$ close to the main neutrality point and reveals the characteristic six-domain structure of the photoresponse, marked by successive sign changes\cite{gabor2011hot, titova2023ultralow}, consistent with the PTE mechanism. This analysis suggests that thermoelectricity is  the dominant mechanism responsible for photovoltage observed in the system.

\begin{figure*}[ht!]
  \centering\includegraphics[width=0.8\linewidth]{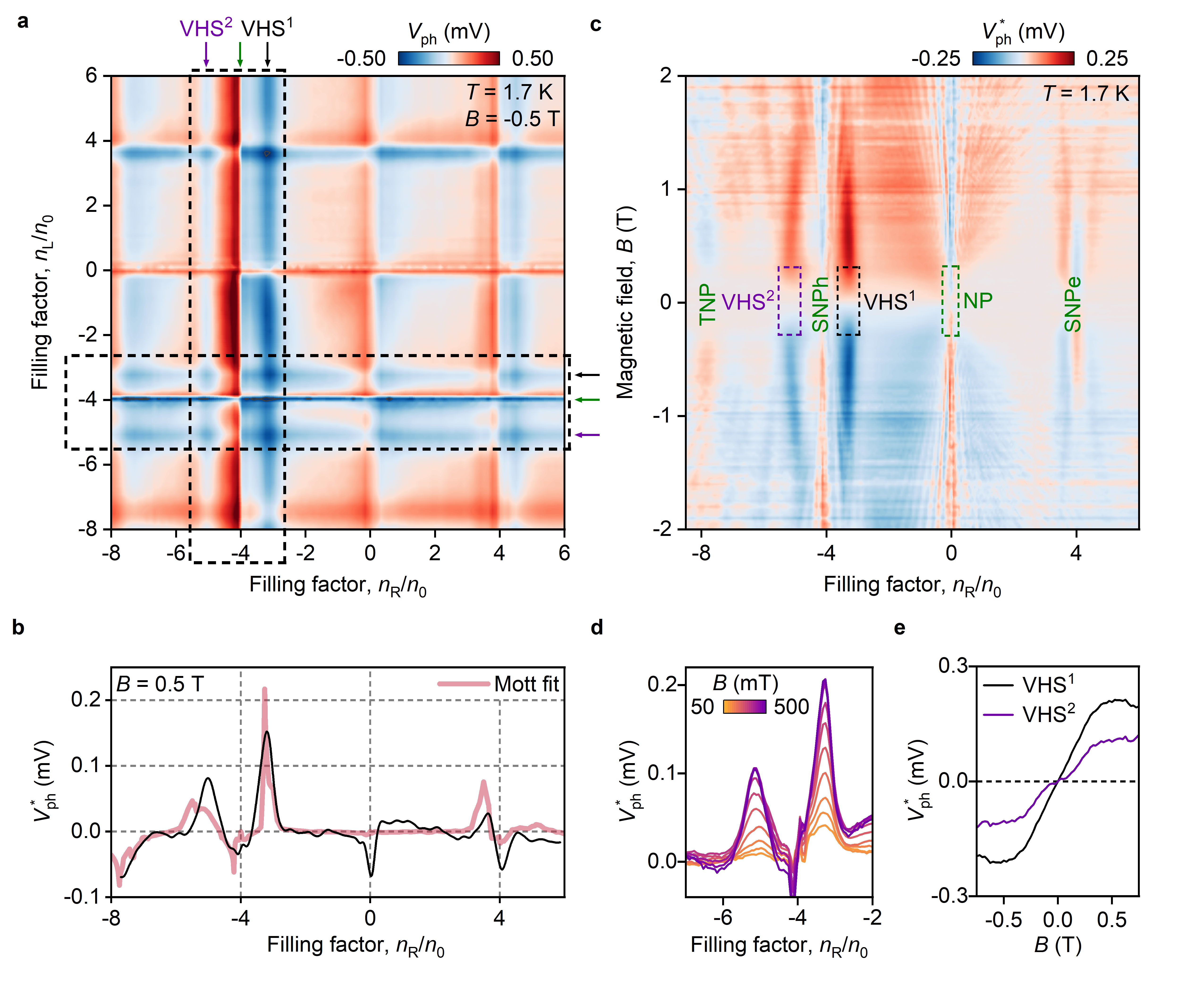}
    \caption{\textbf{THz-driven Nernst effect in MLG/hBN moiré p-n junction.}
    \textbf{a,} Dual-gate map of photovoltage  $V_{\mathrm{ph}}$ at a magnetic field $B$ = -0.5 T. Dashed rectangles highlight the regions of enhanced $V_\mathrm{ph}$ near VHSs.  \textbf{b,} Comparison of antisymmetric component of photovoltage $V^*_\mathrm{ph}$ from experiment with theoretical Mott fit. \textbf{c,} $V^*_{\mathrm{ph}}$ mapped as a function of $n_\mathrm{R}/n_0$ and $B$ at $n_L/n_0 = 3$. To isolate the dependence of $V^*_{\mathrm{ph}}$ on the right-gate doping and to exclude contributions from the rest of the device, individual $V^*_{\mathrm{ph}}(n_\mathrm{R}/n_0)$ traces were shifted such that their value was set to zero at the midpoint between the NP and SNPe
    , where the signal is featureless. \textbf{d,} $V^*_{\mathrm{ph}}$ as a function of  $n_R/n_0$ measured at $B$ ranging from 50 mT (orange) to 500 mT (purple). \textbf{e,} $V^*_{\mathrm{ph}}(B)$ dependence at VHSs. 
    }

	\label{Fig3}
\end{figure*}

The response changes drastically near the SNPs. Figure~\ref{Fig2}a shows the photoresponse across all accessible split-gate configurations. From this point onward, gate voltages are translated to filling factors $n_\mathrm{L(R)}/n_\mathrm{0}$, where $n_\mathrm{L(R)}$ is the carrier density induced by the left (right) split-gate, and $4n_0$ denotes the full filling of a single moir\'e miniband. A pronounced enhancement of the signal is observed near the second neutrality point on the hole side of the moiré miniband (highlighted with a dashed rectangle in Fig.\ref{Fig2}a). Interestingly, the amplification of the normalized photoresponse near the SNPh is highly sensitive to the output power $P$ of the THz source, as shown in Fig. \ref{Fig2}b: the normalized photovoltage $V_{\mathrm{ph}}/(P/P_{\mathrm{0}})$ (where $P_\mathrm{0}$ is the maximum output power)  remains nearly unchanged in regions far from the SNPh, but it increases sharply just below the SNPh. This behavior is indicative of the nonlinear power dependence of the photovoltage in vicinity of the SNPh, since a linear photovoltage–power relationship would yield a constant normalized response. Figure \ref{Fig2}c shows the dependence of the photovoltage as a function of $P/P_{\mathrm{0}}$ and highlights the nonlinear growth of $V_\mathrm{ph}$ near the SNPh in contrast to the linear dependence observed midway between the two neutrality points. Tentatively, we attribute this nonlinear power dependence to superlattice-induced Umklapp $e-e$ scattering, which can strongly influence the thermoelectric response in moiré superlattices~\cite{IvanThermopower}. Finally, we note that the photoresponse in the vicinity of the SNPh remains robust at temperatures up to $T = 40$~K (see Fig.\ref{Fig2}d), reaching the operating range of cost-effective cryocoolers.

To gain a qualitative understanding of the observed enhancement of $V_\mathrm{ph}$, we model the photoresponse using Eq.(\ref{eq2}), with the Seebeck coefficients $S^\mathrm{(L,R)}_{xx}$ given by Eq.(\ref{eq1}), which, in absence of magnetic field, reduces to

\begin{equation}
S^\mathrm{(L,R)}_{xx} 
\sim D(E^\mathrm{(L,R)}) \left( \sigma^{-1}_{xx} \frac{d\sigma_{xx}}{dn}  \right) ^\mathrm{(L,R)}.
\end{equation}
Here, the normalized transconductance $\sigma^{-1}_{xx} \frac{d\sigma_{xx}}{dn}$ was obtained from transport measurements, and the DOS was calculated using the framework described in Jung \textit{et al.}\cite{jung2017moire}. Figure~\ref{Fig2}e shows the results of the modeling $V_{\mathrm{ph}}^{SB}(n_\mathrm{L},n_\mathrm{R})$, which show excellent agreement with the experiment (Figure~\ref{Fig2}a). The responsivity enhancement near the SNPh is evident in Fig.~\ref{Fig2}f, which shows a line cut of $V_{\mathrm{ph}}^{SB}(n_\mathrm{R})$  at $n_\mathrm{L} = -3n_\mathrm{0}$, and is due to the sharp, nonlinear increase in the DOS in this region, associated with reduced $v_\mathrm{F}$ in MS, reinforced by the large $\sigma^{-1}_{xx} \frac{d\sigma_{xx}}{dn}$.



\begin{figure*}[ht!]
  \centering\includegraphics[width=0.8\linewidth]{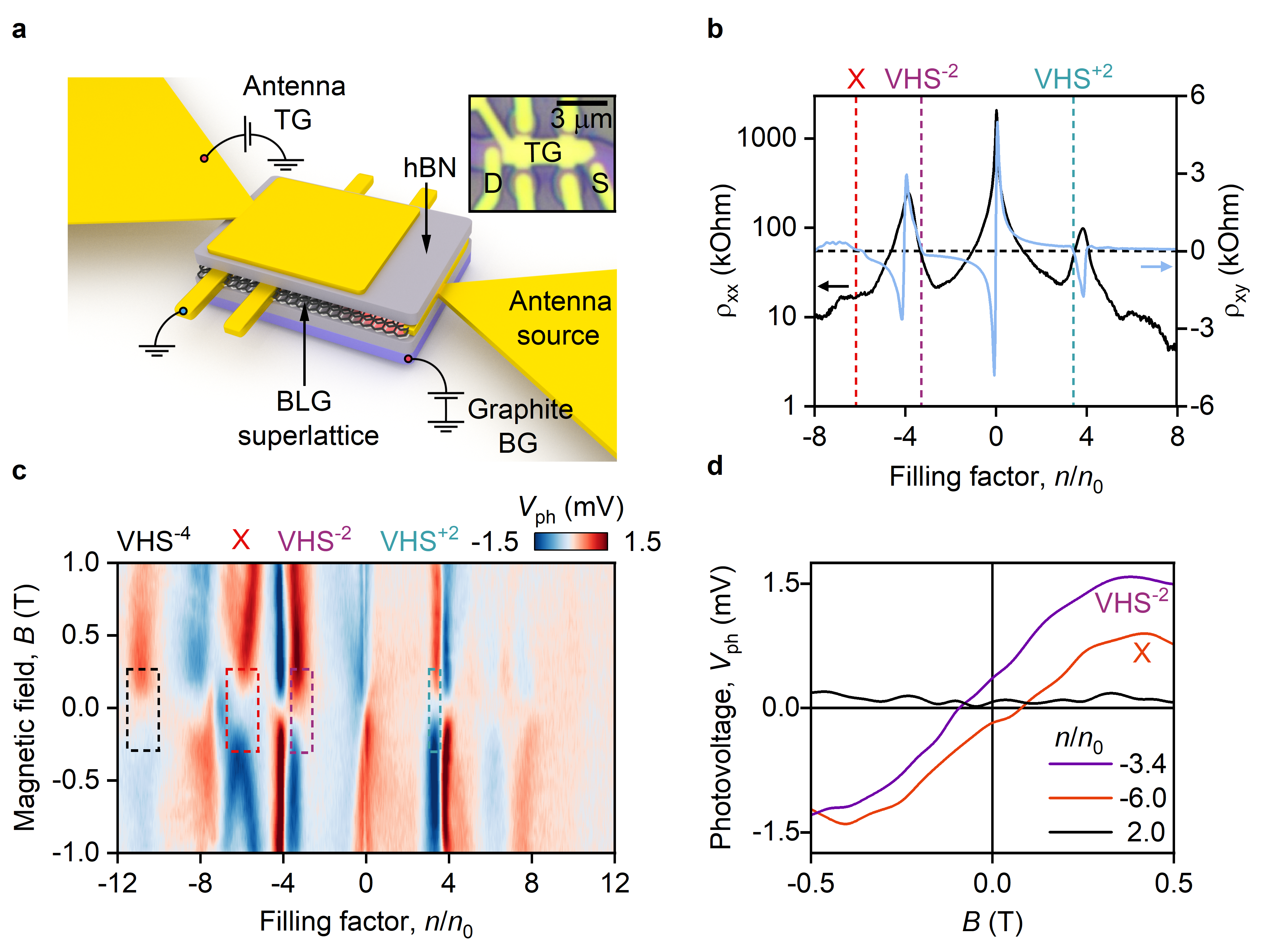}
    \caption{\textbf{THz-driven Nernst effect in the BLG/hBN MS.}
    \textbf{a,} Schematic of the BLG/hBN MS, showing antennas connected to the top gate (TG) and source (S) terminal. Inset: optical photograph of the device. \textbf{b,} Magnetotransport measurements of the longitudinal (black) and Hall (blue) reactivities. Dashed vertical lines mark the position of VHSs. \textbf{c,} $V_\mathrm{ph}$ mapped as a function of magnetic field $B$ and filling factor $n/n_0$ in the BLG/hBN MS. The data exhibits a pronounced antisymmetric response, attributed to the Nernst effect. \textbf{d,} Comparison of $V_{\mathrm{ph}}(B)$ profiles measured at the selected VHSs (purple, orange) and midway between the two neutrality points (black). 
    }

	\label{Fig4}
\end{figure*}

\textbf{THz-driven Nernst effect.} 
To further investigate the PTE effect in graphene MS, we performed photovoltage measurements under finite out-of-plane magnetic field, $B$. 
Figure \ref{Fig3}a shows the dual-gate photovoltage map at $B = -0.5$ T. 
Compared to $V_\mathrm{ph}$ measured at zero field (Fig.~\ref{Fig2}a), two additional features emerge in the $V_{\mathrm{ph}}(n_\mathrm{L},n_\mathrm{R})$ map at finite $B$ at $n_{\mathrm{R,L}} = -3.2 n_{\mathrm{0}}$  and $n_{\mathrm{R,L}} = -5 n_{\mathrm{0}}$ (labeled by arrows). 
These band fillings surprisingly correspond to the VHSs in the DOS which we determined from the Hall effect measurements (Fig.~\ref{Fig1}c). When the magnetic field is reversed to $B = 0.5$ T (Supplementary Information), the signal at these fillings changes sign, suggesting that the observed anomalies are antisymmetric in $B$. 
For this reason, for further analysis, we extracted the antisymmetric component of the photovoltage, $V^{*}_{\mathrm{ph}}$, with respect to $B$ and plotted it against $n_\mathrm{R}$ (Fig. \ref{Fig3}c). At $n_\mathrm{R} = -3.2n_\mathrm{0}$, labeled VHS$^1$, the band filling associated with enhanced $V^{*}_{\mathrm{ph}}$ remains unchanged upon varying $B$. 
In contrast, at $n_\mathrm{R} = -5n_\mathrm{0}$, labeled VHS$^2$, the peak position tends to shift to smaller $n_\mathrm{R}$ with increasing $B$ above 0.5~T \cite{moriya2020emergence}. We further trace peak maxima as a function of $B$ for both VHSs (Fig.~\ref{Fig3}d) and plot its amplitude against $B$ in Fig.~\ref{Fig3}e. The signal grows linearly for $B<0.5~$T, comparable to the magnetic field that can be generated by a permanent magnet. 


The observed enhancement of the photovoltage exhibits two key signatures: symmetry with respect to  $n_{\mathrm{L}}=n_{\mathrm{R}}$ diagonal in the dual-gate map and antisymmetric dependence on $B$. These characteristics suggest that the Nernst effect is the underlying mechanism of photovoltage generation\cite{Checkelsky,cao2016photo,wu2016multiple} near these VHSs. Indeed, in our device geometry, voltage probes $1$ and $2$ are positioned closer to the junction relative to probes $3$ and $4$, giving rise to a transverse temperature gradient that drives the Nernst response $V^{N}_{\mathrm{ph}}$ given by
\begin{equation}
\begin{split}
V_{\mathrm{ph}}^{N} 
= \left[ S_{xy}^\mathrm{(L)} + S_{xy}^\mathrm{(R)} \right] \, \Theta(L),
\end{split}
\end{equation}
where 
$\Theta(L) \equiv \int_{0}^{L} \left\langle \nabla_y T_\mathrm{e}\right\rangle_\mathrm{W} \, dx$ is integrated average (over the channel width $W$) transverse temperature gradient across a split-gate segment of length $L$. From Eq.~(1), the off-diagonal components of the thermoelectric tensor read
\begin{equation}
S_{xy}^{(L,R)} \propto D(E^{(L,R)}_F) \left( \rho_{xx} \frac{d\sigma_{xy}}{dn} + \rho_{xy} \frac{d\sigma_{xx}}{dn} \right)^{(L,R)}
\end{equation}
and are related to the Nernst coefficient as $N = S_{xy} / B$. Here, $\rho_{xx}$ and $\rho_{xy}$ denote the longitudinal and Hall resistivities, respectively.
The physics of the PTE effect near the VHS creates particularly favorable conditions for Nernst enhancement. As the system approaches the VHS, the transverse resistance $\rho_{{xy}}$ approaches zero, effectively suppressing the second term in Eq.~$(5)$ while leaving the first term dominant. Simultaneously, the DOS exhibits a sharp peak, and the Hall conductivity $\sigma_{xy}$ varies rapidly with carrier density. The synergistic interplay of these factors -- enhanced $D(E)$ and large $\frac{d\sigma_{xy}}{dn}$ -- generates a substantial amplification of the Nernst photovoltage. Figure~\ref{Fig3}b demonstrates good qualitative agreement between our theoretical calculations and the experimentally extracted antisymmetric photovoltage, validating this interpretation.

Having established the route to enhance the PTE response in MLG/hBN superlattices, we now turn to a system based on BLG aligned with hBN at a near-zero angle. Owing to its more complex band structure,  BLG/hBN MS hosts multiple VHSs, thereby affording additional tunability of the PTE response. Figure~\ref{Fig4}a presents a schematic of the BLG-based device. In contrast to the split-gate configuration employed in the MLG device, the BLG sample was fabricated in a conventional geometry commonly used for FET-based THz detectors~\cite{dyakonov2002detection,Pseudo-Euler}, namely: the antenna sleeves are connected to the source and the top gate electrodes (inset of Fig.~\ref{Fig4}a: optical micrograph of the device). The antenna-induced high-frequency currents generate substantial local heating near the source contact, establishing a temperature gradient along the device~\cite{bandurin2018dual}. This, combined with contact asymmetry that induces a transverse temperature gradient—similar to that in MLG samples—gives rise to the significant Nernst effect in the system\cite{shilov2024high}. Transport measurements of $\rho_{xx}$ and $\rho_{xy}$ as a function of band filling $n/n_\mathrm{0}$ shown in Fig.~\ref{Fig4}b reveal typical SNPs and indicate the positions of the VHSs within the band structure. 

Figure \ref{Fig4}c shows the map of the as-measured (not-symmetrized) $V_\mathrm{ph}$ as a function of $B$ and $n/n_\mathrm{0}$. As expected, at zero $B$, $V_\mathrm{ph}$ changes sign at the main and secondary neutrality points following the transconductance. When the finite perpendicular $B$ is applied, the total $V_\mathrm{ph}$  develops a pronounced antisymmetric component, strongly enhanced near the VHSs (Fig.\ref{Fig4}c). The X-shaped splitting of the $V_\mathrm{ph}$ near $n/n_\mathrm{0}=-6$ is due to strong topological orbital magnetic moment that leads to valley lifting of the energy bands~\cite{moriya2020emergence}.  Figure~\ref{Fig4}d compares the magnetic field dependence of the photovoltage at $n$ corresponding to the VHSs with that at a filling level midway between two neutrality points and reveals nearly linear growth of $V_\mathrm{ph}$ up to $0.5~$T at VHSs and practically no trend away from them.

\begin{figure}[ht!]
  \centering\includegraphics[width=0.9\linewidth]{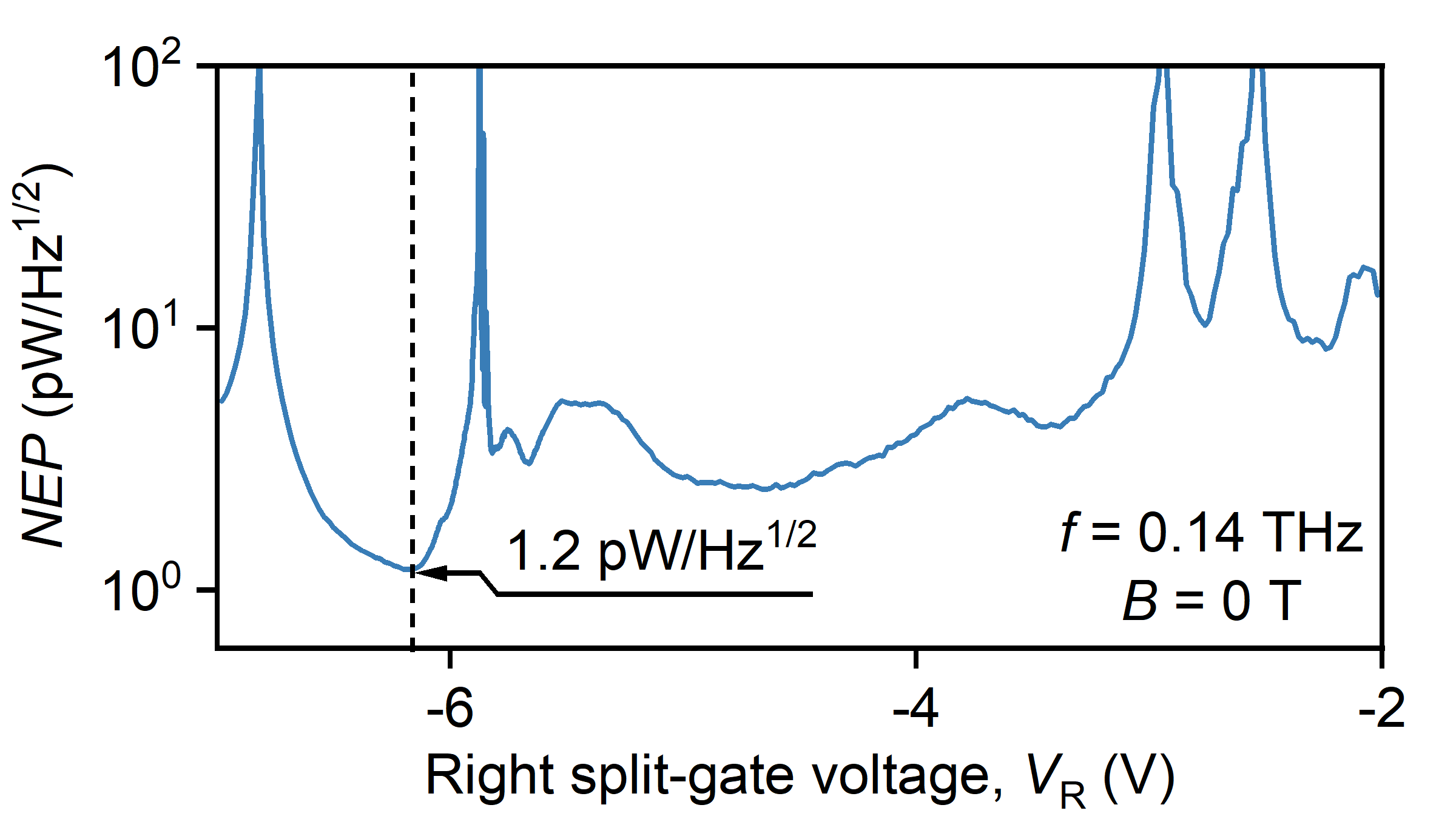}
    \caption{\textbf{Extended Figure. Noise equivalent power at zero \textit{B}.}  NEP as a function of right split-gate voltage $V_\mathrm{R}$ at fixed $V_\mathrm{L}$ at $T=300$ K.
     }
	\label{Fig5}
\end{figure}

\textbf{Detector performance.} Last, we assess the practical performance of our photodetectors. To this end, we characterized their key figures of merit under THz illumination, namely the voltage responsivity $r_\mathrm{V}$ and the noise-equivalent power ($NEP$). By calculating the incident THz power $P_s$ delivered to the antenna area (see Methods), we determined an external responsivity of $r_\mathrm{V} = V_{\mathrm{ph}}/P_s \approx 2.2$~kV/W (i.e., we assumed that the full power impinging on the device is absorbed). The $NEP$, defined as the ratio of voltage noise spectral density $s_\mathrm{V}$ to voltage responsivity ($NEP = s_\mathrm{V}/r_\mathrm{V}$), quantifies the minimum detectable signal in the presence of noise. In our zero-bias detectors, thermal voltage fluctuations dominate following $s_\mathrm{V} = \sqrt{4kTR}$, yielding a minimum NEP of $\sim$1~pW/$\sqrt{\mathrm{Hz}}$ that demonstrates excellent sensitivity for THz applications (See Fig.~\ref{Fig5}. Optimal detector performance was consistently achieved when the Fermi level was electrostatically tuned near the SNP, where moir\'{e}-induced band structure singularities enhance the DOS and maximize the thermoelectric response.

The performance of our singularity-enhanced devices is already competitive with commercially-available SHEBs, which typically reach $r_\mathrm{V} \sim 1$--3~kV/W and $NEP \sim 0.5$--5~pW/$\sqrt{\mathrm{Hz}}$~\cite{klapwijk2017engineering, shurakov2015superconducting, shein2024fundamental}. In our case, the responsivity remains singularity-enhanced up to $\sim$40~K, keeping the device operational well above the $T_\mathrm{c}$ limits of SHEBs. Indeed, weak variation of resistivity in this temperature range ensures $NEP$ to follow the responsivity trend (Supplementary Information). Furthermore, Fig.~\ref{Fig4} shows that applying a modest out-of-plane magnetic field boosts $r_\mathrm{V}$ by a factor of 2--3, reducing the $NEP$ proportionally to the 0.3--0.5~pW/$\sqrt{\mathrm{Hz}}$ range. At present, however, the effective performance is still limited by impedance mismatch between the detector and the antenna, meaning that only a small fraction of incident THz power is absorbed~\cite{kravtsov2025viscous}. This coupling can be substantially improved by integrating high-impedance antenna designs, which are expected to further boost absorption and device responsivity~\cite{inverseMISHA}. Importantly, we note that twist-controlled moiré structures exhibit ultrafast phonon-Umklapp cooling with characteristic times of only a few picoseconds~\cite{TBG-picoseconds}, at least two orders of magnitude faster than the best SHEBs, which typically feature 0.1--1~ns response times, and one order of magnitude faster than in moiré-free graphene~\cite{TBG-picoseconds}. Last, recent advances in the CVD growth of twisted moiré superlattices~\cite{CVDmoire} already provide a practical materials platform, making large-area, pixelated THz detector arrays a realistic prospect.

\section*{Conclusions} 

In summary, we have explored Seebeck and Nernst effects in graphene-based MS under THz irradiation. Our split-gate device configuration enabled precise control over local carrier density, allowing us to probe the PTE response across various band structure features. We observed an enhanced Seebeck photovoltage near the second neutrality point on the hole side of the moiré miniband, which we attribute to a rapid nonlinear increase in the density of states at this filling, associated with a reduced Fermi velocity of the moiré bands. In a finite magnetic field, the photovoltage exhibited a large antisymmetric Nernst component, strongly enhanced near the VHSs. The observed behavior, compared with theoretical modeling based on the Mott formula, was further confirmed experimentally in BLG/hBN MS. Importantly, we demonstrate that the performance of the singularity-enhanced detectors is competitive with commercial SHEBs. Our results establish that moiré-engineered band structures can be exploited to enhance and control thermoelectric responses at THz frequencies, offering new opportunities for sensing applications. It would be interesting to expand such studies to other moiré superlattices such as twisted bilayer graphene where enhanced density of states and interaction effects may drastically enhance THz-driven thermoelectricity and its interplay with other detection mechanisms~\cite{Otteneder2020,Rosh_geom-q,Delgado-Notario2025,Merino2025,ArindhamThermo,DasThermo}.

\section*{Methods} 

\textbf{Sample fabrication.}
Our samples were fabricated from MLG and Bernal-stacked BLG, encapsulated between hBN slabs using a standard dry-transfer technique described elsewhere\cite{castellanos2014deterministic}. One slab of hBN was crystallographically aligned to the graphene sheet by matching their straight edges using an optical microscope equipped with micromanipulators and a high-precision rotation stage. One of the MLG devices employed a bottom graphite split-gate, where the graphite was cut using an atomic force microscope\cite{li2018electrode}. For the second MLG device, as well as the BLG device, narrow-layer graphite strips were attached to the bottom surfaces of the stacks to serve as back gates. The heterostructures were then released onto undoped insulating Si/SiO$_2$ substrates to minimize THz power reflection. Standard electron-beam lithography, reactive ion etching, and thin-film metal deposition were employed to pattern the top gates and contact leads (see Fig.\ref{Fig1}e and Fig.\ref{Fig4}a). For the MLG sample, source and drain electrodes were connected to broadband triangular antennas designed to funnel direct incident THz radiation into the channel\cite{titova2023ultralow}. In contrast, the BLG sample had the conventional source-gate antenna coupling\cite{bandurin2018resonant}.

\textbf{Photovoltage measurements.}
The samples were placed in the chamber of a Quantum Design OptiCool magneto-optical variable-temperature cryostat. The incident radiation was linearly polarized along the antenna axis, with its power was controlled by an attenuator. To ensure symmetric illumination, we measured photovoltage maps as a function of the lens x–y position, verifying that the signal changed sign between pn and np doping configurations, as required by symmetry.

DC voltages were independently applied to the three gates using Keithley SM2614B and SRS SIM928, and leakage currents were monitored. All photovoltage data were collected with the zero-bias current using standard lock-in techniques with several SR860 lock-in amplifiers referenced to the laser modulation frequency (37 Hz). To derive the peak photovoltage ($V_{\mathrm{ph}}$) from the root-mean-square (RMS) voltage measured by the lock-in amplifier ($V_{\mathrm{ph}}^{\mathrm{lock-in}}$), we applied the conversion factor $V_{\mathrm{ph}} = (\pi/2) \sqrt{2}V_{\mathrm{ph}}^{\mathrm{lock-in}}$. This factor accounts for the Fourier expansion of square wave used to modulate the THz source, and the conversion from the lock-in's RMS measurement to a peak-to-peak voltage amplitude. For transport measurements, we applied a small AC current (200 nA RMS) using a voltage-controlled current source (SRS CS580) at a lock-in frequency of 11 Hz.

\textbf{Responsivity evaluation.} The responsivity evaluation was based on the pre-calibrated~\cite{titova2023ultralow} source parameters: source power $P_0$ = 16.4 mW and a beam divergence angle of 20°. The optical path consisted of a TPX lens system  and an attenuator with a transmission of 0.03$\%$. We also accounted for the 19$\%$ absorption of THz radiation by the cryostat windows, which we measured separately~\cite{Titova2025}. Considering all the above factors—including the attenuation, cryostat window absorption, and the geometry of the optical setup—the resulting power incident on the antenna was calculated to be 340 nW. The responsivity measurements were peformed on monolayer graphene sample with graphite split gate.



\vspace{1em}







\bibliography{Bibliography.bib}

\newpage
\setcounter{figure}{0}
\renewcommand{\thesection}{}
\renewcommand{\thesubsection}{S\arabic{subsection}}
\renewcommand{\theequation} {S\arabic{equation}}
\renewcommand{\thefigure} {S\arabic{figure}}
\renewcommand{\thetable} {S\arabic{table}}

\end{document}